\documentclass[letterpaper]{article}

\usepackage[T1]{fontenc}
\usepackage[utf8]{inputenc}
\usepackage{geometry}
\usepackage{setspace}
\usepackage{float}
\usepackage{placeins}
\usepackage{achemso}

\usepackage{graphicx}
\usepackage{float}
\usepackage{amssymb,amsmath}
\usepackage{rotating}
\usepackage{longtable}
\usepackage{subcaption}
\usepackage{booktabs}
\usepackage[version=4]{mhchem}
\usepackage{siunitx}
\usepackage{pgfplots}
\pgfplotsset{compat=1.18}
\usepackage{xcolor}
\usepackage[hidelinks]{hyperref}

\newfloat{scheme}{htbp}{los}
\floatname{scheme}{Scheme}
\newfloat{chart}{htbp}{loc}
\floatname{chart}{Chart}
\newfloat{graph}{htbp}{loh}

\def\*{\phantom{0}}
\makeatletter
\providecommand\barcirc{\mathpalette\@barred\circ}
\def\@barred#1#2{\ooalign{\hfil$#1-$\hfil\cr\hfil$#1#2$\hfil\cr}}

\makeatother

\usepackage{authblk}

\author[1,2,3,4]{Bouthe\"ina Kerkeni\thanks{Corresponding author. Email: boutheina.kerkeni@obspm.fr}}
\author[5]{Ghofrane Ouerfelli}
\author[4]{Nicole Feautrier}
 \author[4]{Christian Balan\c ca}

\affil[1]{De Vinci Higher Education, De Vinci Research Center, 92916 Paris, France}
\affil[2]{Institut Supérieur des Arts Multimédia de la Manouba, Université de la Manouba, 2010 la Manouba, Tunisia}
\affil[3]{Facult\'e des Sciences de Tunis, Laboratoire de Physique de la Mati\`ere Condens\'ee, Universit\'e Tunis El Manar, 2092 Tunis, Tunisia}
\affil[4]{Observatoire de Paris, Universit\'e PSL, CNRS-UMR 8112, LUX, F-92195 Meudon, France}
\affil[5]{Department of Physics, College of Khurma University, Taif University, P.O. Box 11099, Taif, 21944, Saudi Arabia}
 
\title{Reaction Mechanisms and Kinetics of CN and CCH with H$_2$CS: Implications for Interstellar Sulfur Chemistry}
\date{}

\begin{document}

\maketitle

\begin{abstract}
We report an \textit{ab initio} and master-equation investigation of the gas-phase reactions of thioformaldehyde (H$_2$CS) with CN and CCH radicals, motivated by the recent detection of the S-containing species HCSCN and HCSCCH in cold interstellar environments. Structures and frequencies were obtained at the DSD-PBEP86/aug-cc-pVTZ level, with energetics refined by CCSD(T)-F12a calculations and kinetics treated using an energy-grained master equation.

The CN + H$_2$CS reaction proceeds through orientation-dependent entrance channels. Two barrierless addition pathways lead to a connected multi-well network that preferentially forms cyano thioformaldehyde, HCSCN + H, whereas abstraction-type channels leading to HNC + HCS or HCN + HCS contribute only marginally. The calculated kinetics indicate capture-controlled low-temperature reactivity and a strong preference for HCSCN formation, suggesting that this reaction should be considered in astrochemical models of cold clouds.

For CCH + H$_2$CS, barrierless capture gives access to two connected entrance adducts. Although the cyclic product is the most exothermic channel, its formation is kinetically hindered by a high-lying late transition state. The flux is shared between propynethial formation, HCSCCH + H, and the HCCH + HCS channel, with the latter being favored. These results show that subtle differences in radical structure, bonding preferences, and entrance-channel topology strongly affect product branching in S-containing radical–molecule reactions. The computed mechanisms and rate coefficients provide useful input for astrochemical models of sulfur chemistry in cold molecular clouds and for interpreting recent molecular detections in sources such as TMC-1.

\end{abstract}

\section*{Keywords}
Sulfur chemistry; interstellar medium; thioformaldehyde; CN radical; CCH radical; HCSCN; HCSCCH; \textit{ab initio} calculations; master equation; reaction kinetics

\section*{Abbreviations}
ISM, interstellar medium; PES, potential energy surface; ZPE, zero-point energy; IRC, intrinsic reaction coordinate; MESS, master-equation solver for multi-energy well reactions; DFT, density functional theory; TMC-1, Taurus Molecular Cloud 1

\section{Introduction}

Sulfur chemistry in the interstellar medium (ISM) remains one of the major
unresolved problems in astrochemistry. Although sulfur has a relatively high cosmic abundance, with a solar reference abundance of 
$\log\epsilon({\rm S}) = 7.12 \pm 0.03$ 
($S/H \simeq 1.3\times10^{-5}$), its observable gas-phase abundance in cold and dense molecular environments is often found to be strongly depleted relative to this cosmic value \citep{Asplund2009}.
 The elemental abundance of sulfur in
molecular clouds remains uncertain by several orders of magnitude, and no
unified scheme currently accounts for the observed abundances of
sulfur-bearing species across different interstellar environments
\cite{Fuente2023}.

Several studies have therefore investigated the nature of the missing sulfur
reservoir. Chemical-network modeling of dark clouds has suggested that the
observed sulfur-bearing molecules account for only a small fraction of the
expected elemental sulfur budget, with possible reservoirs including atomic
sulfur, HS/H$_2$S-related ice chemistry, and other grain-surface products
\cite{Vidal2017}. In this context, grain-surface models have proposed that a
significant fraction of the missing sulfur may be stored in organo-sulfur
species trapped on dust grains \cite{Laas2019}. Observational and chemical
studies of thioformaldehyde and related sulfur species further point to icy
mantles and S-bearing molecules such as H$_2$S, OCS, SO$_2$, H$_2$S$_2$, CS$_2$,
or S$_8$ as possible reservoirs or tracers of sulfur chemistry
\cite{Esplugues2022}.

In this context, the cold dark cloud TMC-1 has emerged as a key benchmark for probing sulfur chemistry under low-temperature conditions. The discovery of cyano thioformaldehyde (HCSCN) and propynethial (HCSCCH) in TMC-1 provided direct motivation for investigating cyanide and ethynyl chemistry involving thioformaldehyde in cold clouds \cite{Cernicharo2021}. More broadly, spectral surveys of TMC-1, including QUIJOTE, have revealed an unexpectedly rich molecular inventory in this source, encompassing carbon-chain molecules, radicals, complex organic species, and S-containing compounds such as NCS, HCCS, H$_2$CCS, H$_2$CCCS, C$_4$S, C$_5$S, HCCS$^+$, NC$_3$S, HC$_3$S, and HS$_2$ \cite{Kaifu2004,Cernicharo2021b,Cernicharo2021d,Cernicharo2022,Cabezas2022,Cernicharo2024b,Esplugues2025,Fuentetaja2022,Cabezas2024,Fuentetaja2025}. Earlier studies of sulfur chemistry in cold dark clouds have provided important constraints for interpreting these observations \cite{Agundez2019}.

These observations reveal abundance ratios that cannot be straightforwardly
explained by simple analogies with oxygen chemistry. HCSCN is observed to be
significantly more abundant than its oxygen analogue HCOCN, whereas HCSCCH is
less abundant than propynal (HCOCCH), leading to an inverted HCSCCH/HCSCN ratio
compared with related oxygen-bearing cyanide–ethynyl pairs
\cite{Cernicharo2021,Rodriguez2021,Agundez2025,Sanz2025}. Such contrasts suggest that fast radical–neutral reactions may play a key role
in sulfur chemistry in cold dark clouds, while also highlighting the need for
accurate reaction kinetics and product branching ratios
\citep{ACSESC2024,NatComm2022_Sulfur,NatComm2025_Sulfur,Yang2024}.

Radical–neutral reactions involving CN and CCH are generally considered
efficient under interstellar conditions owing to their often barrierless or
weakly hindered entrance channels, in line with the general importance of rapid
neutral–neutral reactions at low temperatures \cite{Smith2004}. 

The CN + H$_2$CO reaction has long served as a prototypical benchmark system and has
been extensively investigated both experimentally and theoretically. Subsequent
kinetic and theoretical studies demonstrated that, despite efficient capture,
the reaction predominantly yields HCN + HCO, whereas the formation of formyl
cyanide (HCOCN) is kinetically disfavored
\cite{Chang95,Vuitton2012_CN_H2CO,Tonolo20,West2023_CN_H2CO}. Related
low-temperature CN-radical reactivity has also been explored for aromatic
nitrogen heterocycles relevant to cold interstellar environments
\cite{Heitkamper2022_PyridineCN}.

The comparison with the CN + H$_2$CO benchmark is particularly instructive.
West et al. \citep{West2023_CN_H2CO} showed that direct low-temperature measurements, combined with
MESMER calculations constrained by experiment, lead to rate coefficients in the
$10^{-11}$ cm$^3$ s$^{-1}$ range and to HCN + HCO as the dominant product,
whereas the HCOCN-forming channel is hindered by a large activation barrier.
Their work also demonstrates that small changes in entrance-region barriers and
transition-state parameters can strongly affect the predicted low-temperature
rate coefficients. This point is important for the sulfur analogue, because
substitution of O by the more polarizable S atom is expected to modify
long-range interactions, entrance-channel topology, and the connectivity of the
potential energy surface.

In astrochemical networks, such reactions are usually represented by simplified
global rate coefficients \cite{Millar1997}. These implementations do not
explicitly resolve the orientation dependence of the entrance region or the
subsequent multi-well dynamics. This limitation is particularly relevant for
reactions involving polar and highly polarizable molecules, for which
long-range electrostatic and dispersion interactions may influence the capture
step and product branching. An explicit entrance-resolved master-equation
treatment is therefore needed to assess whether different approach geometries
lead to distinct addition, abstraction, or product-forming pathways.

Attention has recently turned to the sulfur analogue reactions involving
thioformaldehyde (H$_2$CS). A recent high-level theoretical study by
Alessandrini and co-workers \cite{Alessandrini2025} investigated the
CN + H$_2$CS reaction and identified HCSCN + H and HCN + HCS as energetically
accessible product channels. However, the role of orientation-dependent
entrance capture, the separation between addition and abstraction manifolds,
and the corresponding product branching remain important kinetic questions.
Moreover, no comparable potential-energy-surface and temperature-dependent
master-equation study appears to be available for the CCH + H$_2$CS reaction,
although this reaction has been proposed as a plausible gas-phase route to
HCSCCH in TMC-1. This motivates the present combined quantum-chemical and
master-equation investigation.

For such radical–molecule reactions, subtle features of the potential energy surface—such as the
connectivity between entrance channels and multiwell reaction networks—can
strongly influence product selectivity.

In this work, we present a unified \textit{ab initio} and kinetic investigation
of the CN + H$_2$CS and CCH + H$_2$CS reactions.
Stationary points are characterized at the DSD-PBEP86/aug-cc-pVTZ level and
refined using CCSD(T)-F12a single-point energies including zero-point energy
corrections.
The resulting multi-well reaction networks are treated using an energy-grained
master-equation approach implemented in \textsc{MESS}
\citep{Georgievskii2013_MESS}, allowing us to derive temperature-dependent
phenomenological rate coefficients and product branching ratios in the
low-pressure regime over the numerically converged temperature range accessible
in the present calculations.

Section~2 describes the theoretical methods employed in this study, including
the electronic-structure calculations and the master-equation kinetic treatment.
Section~3 presents the computed potential energy surfaces, reaction mechanisms, rate coefficients, product branching ratios, and analytical
Kooij representations. Section~4 discusses the astrochemical implications of
the computed kinetics for the formation of HCSCN and HCSCCH in cold
interstellar environments, and Section~5 summarizes the main conclusions.
 
\section{Computational Methods}

\subsection{Quantum chemical calculations}

The reaction mechanisms of the CN and CCH radicals with thioformaldehyde
(H$_2$CS) were investigated using high-level quantum chemical calculations.
The potential energy surfaces (PESs) were initially explored at the
DSD-PBEP86/aug-cc-pVTZ level in order to locate all relevant stationary points,
including reactants, intermediates, transition states, and products
\citep{Grimme2011,Kozuch2013,Santra2019}.

Preliminary MP2 geometry optimizations and relaxed entrance-channel scans were
also tested for the CN + H$_2$CS system. However, these calculations were found
to be sensitive to the underlying open-shell Hartree--Fock reference, especially
along dissociation and long-range approach coordinates. In some regions of the
scan, different self-consistent-field solutions of the same spatial symmetry
led to different MP2 energies and to discontinuities or state-switching
artifacts in the entrance potential. This behavior is consistent with the known
dependence of MP2 on the quality and stability of the Hartree--Fock reference
for open-shell or near-degenerate systems. We therefore did not retain MP2 as
the primary geometry-optimization method. Instead, the double-hybrid
DSD-PBEP86 functional was used to provide a more robust correlated reference for
the characterization of stationary points and entrance geometries, while final
energetics were refined by CCSD(T)-F12a single-point calculations.

Using this protocol, all geometries were fully optimized, and harmonic vibrational frequency calculations were performed to characterize the nature of each stationary
point and to obtain zero-point energy (ZPE) corrections.
Transition states were identified by the presence of a single imaginary
frequency and were further validated through intrinsic reaction coordinate
(IRC) calculations to confirm their connectivity to the appropriate minima.
All DSD-PBEP86 calculations were carried out using the \textsc{Gaussian} suite
of programs \cite{gauss}.

To refine the energetic description, single-point energy calculations were
performed for all optimized structures using the explicitly correlated
coupled-cluster method CCSD(T)-F12a, as implemented in the \textsc{Molpro}
program package \cite{MOLPRO,Molpro2020,knowles:93,knowles:00,Knizia2009,Adler2007}.
The aug-cc-pVTZ (AVTZ) basis set \citep{dunning:89,woon:93,Kendall1992,Peterson2008} was employed,
together with the corresponding auxiliary basis sets required for
density-fitting and resolution-of-the-identity techniques.
 
The final relative energies reported in this work correspond to
CCSD(T)-F12a/AVTZ electronic energies augmented by ZPE corrections obtained
at the DSD-PBEP86/AVTZ level.

\subsection{Master-equation kinetics}
The kinetics of the reaction networks were investigated using an
energy-grained master-equation approach as implemented in the
\textsc{MESS} code (Master Equation Solver for Multiwell reactions)
\citep{Georgievskii2013_MESS}.
 
Microcanonical rate constants for elementary steps involving transition states
were calculated within Rice--Ramsperger--Kassel--Marcus (RRKM) theory
\citep{Marcus1952,Jasper2007}. RRKM-based microcanonical rate constants are
also used in related energy-grained master-equation implementations, including
MESMER \citep{Glowacki2012_MESMER}.

Barrierless association processes were treated using capture theory, with the
long-range interaction represented by an effective attractive potential of the
form $-C_6/R^6$. The corresponding $C_6$ parameters were obtained from relaxed
entrance-channel scans and used to describe the barrierless formation of
entrance complexes at long range
\citep{Georgievskii2003,Heitkamper2022_PyridineCN}.
Quantum tunneling effects were included for transition states involving
hydrogen-atom transfer or abstraction using a one-dimensional Eckart barrier
model \citep{Eckart1930}, a commonly used approximate tunneling correction in chemical rate calculations \citep{Lohle2018_JCTC_RateConstants}. All other
transition states were treated within the classical RRKM framework.
\paragraph{Energy graining}
The master equation was solved on an energy grid whose step size scales with
temperature according to
\[
\Delta E = \alpha k_{\mathrm B}T,
\]
where $\alpha$ is the dimensionless parameter specified by the
\texttt{EnergyStepOverTemperature} keyword in MESS. In the present calculations,
$\alpha$ was set to 0.2. This value provides a fine discretization relative to
the thermal energy scale while keeping the computational cost moderate.
 
\paragraph{Bath gas and energy-transfer model}
Master-equation calculations were performed using He as an inert reference bath
gas, together with a Lennard--Jones collision-frequency model and an
exponential-down model for collisional energy transfer. Unless otherwise stated,
all simulations were carried out at $P=10^{-7}$~atm, which was used as a
low-pressure limit for the phenomenological master-equation treatment. This
choice follows recent astrochemical master-equation studies of CN-radical
reactions performed with the same \textsc{MESS} framework at $10^{-7}$~atm
\citep{Heitkamper2022_PyridineCN}.

Additional test calculations performed at lower pressures did not lead to
significant changes in the product-forming rate coefficients or branching
fractions, indicating that the reported kinetics are effectively in the
low-pressure regime and are not controlled by collisional stabilization.
Accordingly, the reported product branching fractions are mainly governed by
entrance capture, unimolecular redistribution, and competition between
product-forming channels within the reactive network. Temperature-dependent
rate coefficients and product branching ratios were then analyzed over the
lowest numerically stable temperature range accessible in the present
calculations.

\section{Results and Discussion}

Here, ``submerged'' denotes a transition state lying below the energy of the
separated reactants at the CCSD(T)-F12a+ZPE//DSD-PBEP86 level, whereas ``emerged''
indicates a barrier above reactants.
We present the computed potential energy surfaces and reaction mechanisms for
the gas-phase reactions of thioformaldehyde (H$_2$CS) with the CN and CCH radicals,
with emphasis on the orientation-dependent entrance channels and their
connection to competing product pathways relevant to interstellar sulfur
chemistry.

\subsection{Reaction Networks and Energetics}
\begin{figure}[H]
\begin{center}
\includegraphics[width=15cm]{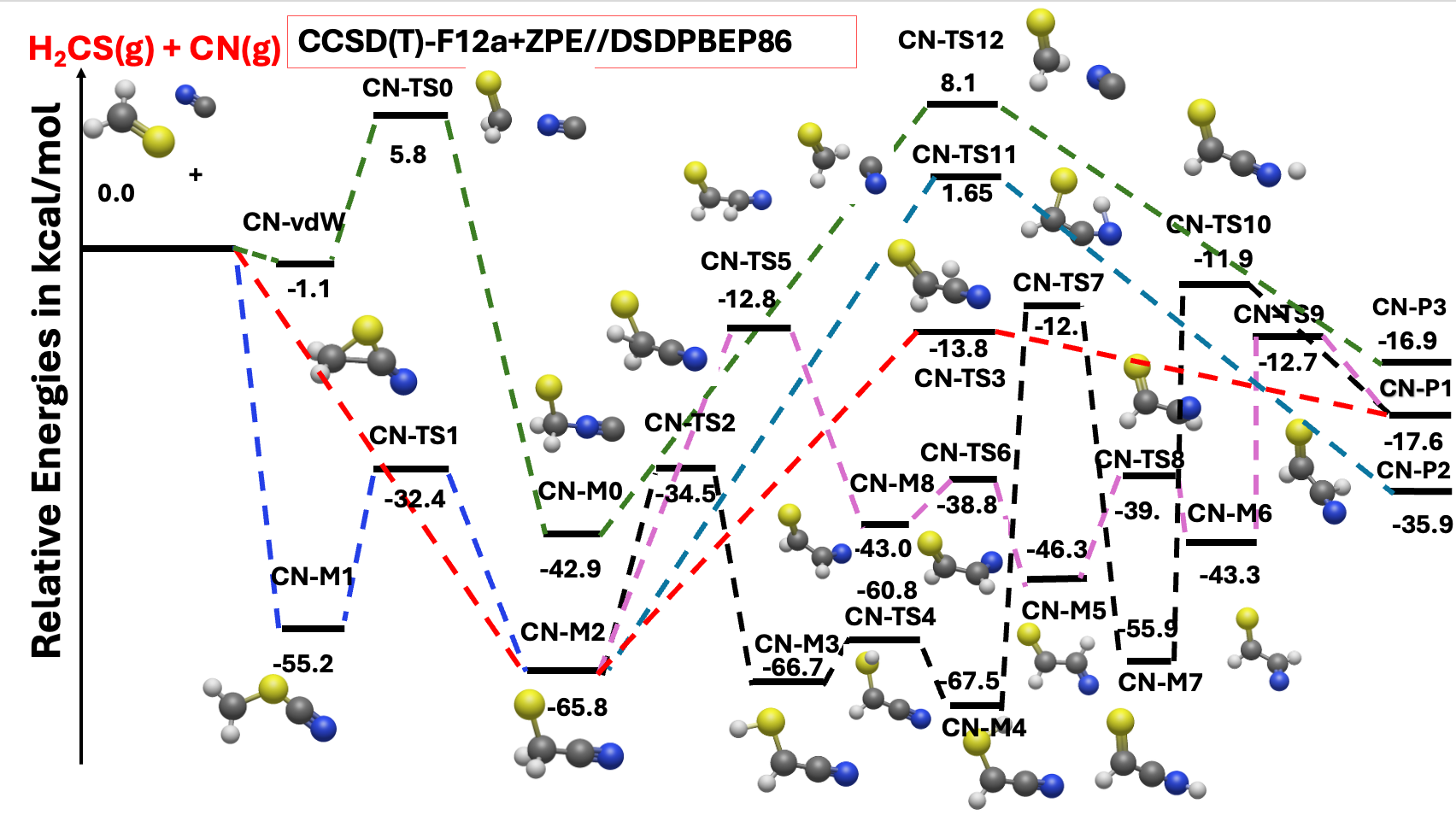}
\caption{Reaction mechanism for the CN + H$_2$CS system computed at the
CCSD(T)-F12a+ZPE//DSD-PBEP86/aug-cc-pVTZ level. Relative energies are given
in kcal mol$^{-1}$ with respect to the separated reactants.}
\label{Figure:1a}
\end{center}
\end{figure}

\begin{figure}[H]
\begin{center}
\includegraphics[width=15cm]{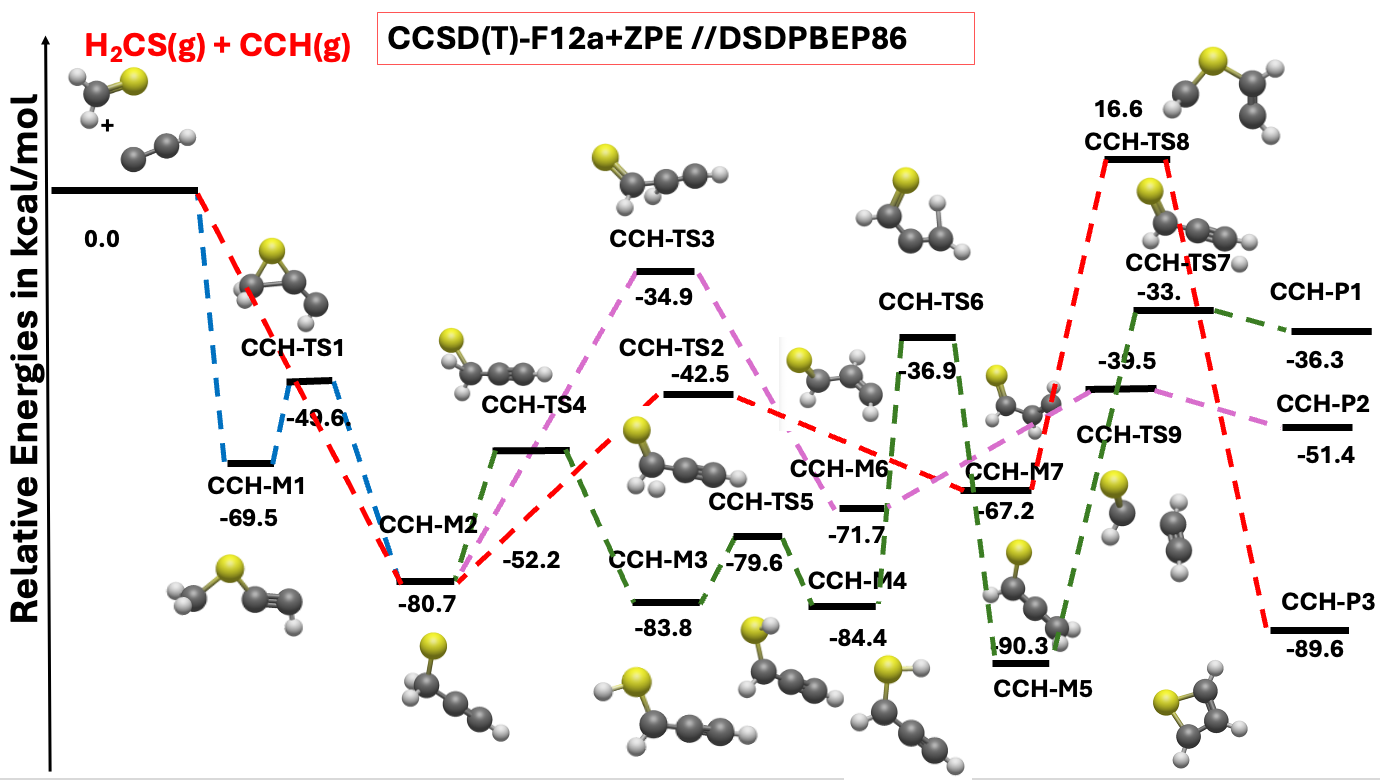}
\caption{Reaction mechanism for the CCH + H$_2$CS system computed at the
CCSD(T)-F12a+ZPE//DSD-PBEP86/aug-cc-pVTZ level. Relative energies are given
in kcal mol$^{-1}$ with respect to the separated reactants.}
\label{Figure:1b}
\end{center}
\end{figure}

 For clarity, the complete CN and CCH reaction networks are summarized below using a consistent CN–M$_x$/TS$_x$/P$_x$ and CCH–M$_x$/TS$_x$/P$_x$ notation. Here, M$_x$ denotes stationary-point minima on the potential energy surface, including entrance complexes and reaction intermediates; TS$_x$ denotes first-order saddle points connecting adjacent minima, as verified by intrinsic reaction coordinate calculations; and P$_x$ denotes product channels or product wells, including asymptotic separated products where applicable. Figures~\ref{Figure:1a} and~\ref{Figure:1b} display the reaction pathways derived for the CN + H$_2$CS and CCH + H$_2$CS reactions, respectively. 
 
\subsection{CN + H$_2$CS reaction}

For CN + H$_2$CS, the landscape is strongly orientation dependent. The S-bound
and C/H-side approaches lead barrierlessly to CN--M1 and CN--M2, respectively.
In contrast, only the N-side approach involves the formation of a long-range
entrance complex, CN--vdW, which is followed by passage through the shallow
entrance transition state CN--TS0 to form CN--M0.

The CN--M1 and CN--M2 entrances rapidly merge, since CN--M1 rearranges to
CN--M2 via CN--TS1, and both feed a common multi-well addition network.

From CN–M2, the system branches into two main product channels. The dominant pathway leads to cyano thioformaldehyde formation, HCSCN + H (CN–P1), through several competing rearrangement sequences, involving either the CN–M3/CN–M4/CN–M7 chain, the CN–M8/CN–M5/CN–M6 chain, or a more direct route via CN–TS3. A secondary channel forms HCN + HCS (CN–P2) via CN–TS11, which lies 1.65~kcal~mol$^{-1}$ above the separated reactants. In contrast, the HNC + HCS channel (CN–P3) originates exclusively from the abstraction-type intermediate CN–M0, which is accessed from the van der Waals complex via the shallow entrance transition state CN–TS0 and subsequently proceeds through CN–TS12.

At the DSD-PBEP86 level, the entrance region displays two additional shallow
features: a weak van der Waals complex along the C-side approach and a very
shallow entrance transition state, CN--TS00, along the pathway leading to
CN--M2. Neither feature is retained at the CCSD(T)-F12a level.

At the DSD-PBEP86/aug-cc-pVTZ level, several entrance-channel transition states
are found to be only weakly emerged, typically lying $\sim$1--3~kcal~mol$^{-1}$
above the reactants, such as CN--TS0 and CN--TS12. When refined at the
CCSD(T)-F12a+ZPE//DSD-PBEP86 level, these barriers increase to approximately
6--9~kcal~mol$^{-1}$, reflecting the sensitivity of shallow entrance regions
to the electronic-structure treatment.

The present reaction network is therefore not intended to be a one-to-one
reproduction of the pathway set reported by Alessandrini et al.
Differences in the number and connectivity of located pathways may arise from
the level of theory used for geometry optimization, the treatment of the
open-shell entrance region, and the strategy adopted for exploring the potential
energy surface. In preliminary tests, MP2 optimizations and scans suggested
additional apparent pathways in some entrance regions. However, these features
were sensitive to the underlying Hartree--Fock reference and did not persist as
stable minima or transition states upon DSD-PBEP86 optimization and
CCSD(T)-F12a energy refinement. A similar sensitivity of shallow entrance
barriers to the level of theory has been reported for the CN + H$_2$CS system
\citep{Alessandrini2025}.

\[
\text{CN + H}_2\text{CS}  \rightarrow
\left\{
\begin{array}{l}
  \text{CN--M2} \\[6pt]
\text{CN--M1} \rightarrow \text{CN--TS1} \rightarrow \text{CN--M2} \\[6pt]
\text{CN--vdW}\rightarrow \text{CN--TS0} \rightarrow \text{CN--M0}  \rightarrow \text{CN--TS12} \rightarrow \text{CN--P3}
\end{array}
\right.
\]
\[
\text{CN--M2} \rightarrow
\left\{
\begin{array}{l}
\text{CN--TS2} \rightarrow \text{CN--M3}
\rightarrow \text{CN--TS4} \rightarrow \text{CN--M4}
\rightarrow \text{CN--TS7} \rightarrow \text{CN--M7}
\rightarrow \text{CN--TS10} \rightarrow \text{CN--P1} \\[6pt]

\text{CN--TS5} \rightarrow \text{CN--M8}
\rightarrow \text{CN--TS6} \rightarrow \text{CN--M5}
\rightarrow \text{CN--TS8} \rightarrow \text{CN--M6}
\rightarrow \text{CN--TS9} \rightarrow \text{CN--P1} \\[6pt]
\text{CN--TS3} \rightarrow \text{CN--P1} \\[6pt]
\text{CN--TS11} \rightarrow \text{CN--P2}
\end{array}
\right.
\]

\begin{table}[htbp]
\centering
\caption{
Relative energies (in kcal~mol$^{-1}$) of stationary points involved in the
CN + H$_2$CS reaction.
Energies are given relative to the separated reactants CN + H$_2$CS.
The first column corresponds to CCSD(T)-F12a+ZPE//DSD-PBEP86, and the second to
DSD-PBEP86+ZPE.
}
\label{tab:cn_energetics}
\begin{tabular}{l c c}
\hline\hline
Species &
$\Delta E_{\mathrm{CCSD(T)\mbox{-}F12a+ZPE//DSD-PBEP86}}$ &
$\Delta E_{\mathrm{DSD-PBEP86+ZPE}}$ \\
\hline
CN + H$_2$CS (reactants) & 0.00  & 0.00 \\
CN--vdW                 & -1.1 & -0.19 \\
CN--TS0                 & 5.8   &2.86 \\
CN--TS00                & --  & 0.21 \\
CN--TS1                 & -32.4 & -38.02 \\
CN--TS2                 & -34.5 & -42.25 \\
CN--TS3                 & -13.8 & -18.78 \\
CN--TS4                 & -60.8 & -67.66 \\
CN--TS5                 & -12.8 & -18.11\\
CN--TS6                 & -38.8 & -44.23 \\
CN--TS7                 & -12.03 & -21.63 \\
CN--TS8                 & -39.04 & -40.82 \\
CN--TS9                 & -12.7 & -18.11 \\
CN--TS10                & -11.9 & -15.31 \\
CN--TS12                & 8.1   & 0.74\\
CN--TS11                & 1.65  & -3.53 \\
CN--M0                  & -42.9 & -48.04 \\
CN--M1                  & -55.2 & -63.03 \\
CN--M2                  & -65.8 & -75.62\\
CN--M3                  & -66.7 & -73.89 \\
CN--M4                  & -67.5 & -74.63 \\
CN--M5                  & -46.3 & -49.3 \\
CN--M6                  & -43.3 & -46.03 \\
CN--M7                  & -55.9 & -65.58\\
CN--M8                  & -42.9 & -48.03\\
CN--P1 (HCSCN + H)      & -17.6 & -26.14 \\
CN--P2 (HCN + HCS)      & -35.9 & -40.97 \\
CN--P3 (HNC + HCS)      & -16.9 & -21.28 \\
\hline\hline
\end{tabular}
\end{table}

\subsection{CCH + H$_2$CS reaction}

For CCH + H$_2$CS, the formation of chemically bound entrance adducts is also strongly orientation dependent. In the CCSD(T)-F12a description, the entrance region leads directly to two bound adducts, CCH--M1 and the markedly deeper CCH--M2. In contrast to the DSD-PBEP86 profile, no shallow entrance van der Waals complex or weakly emerged entrance transition state is retained.

The CCH--M2 intermediate constitutes the principal branching point of the
reaction network.
From this deep well, the system efficiently evolves toward the more
exothermic HCCH + HCS channel (CCH--P2) through the sequence
CCH--TS3/CCH--M6/CCH--TS9.
A competing pathway leads to the formation of propynethial
(HCSCCH + H; CCH--P1) via the multi-step rearrangement sequence
CCH--TS4/CCH--TS5/CCH--TS6/CCH--TS7.

\[
\text{CCH + H}_2\text{CS}   \rightarrow
\left\{
\begin{array}{l}
 \text{CCH--M2} \\[6pt]
\text{CCH--M1} \rightarrow \text{CCH--TS1} \rightarrow \text{CCH--M2}
\end{array}
\right.
\]
\[
\text{CCH--M2} \rightarrow
\left\{
\begin{array}{l}
\text{CCH--TS3} \rightarrow \text{CCH--M6}
\rightarrow \text{CCH--TS9} \rightarrow \text{CCH--P2} \\[6pt]

\text{CCH--TS4} \rightarrow \text{CCH--M3}
\rightarrow \text{CCH--TS5} \rightarrow \text{CCH--M4} \\
\qquad\rightarrow \text{CCH--TS6} \rightarrow \text{CCH--M5}
\rightarrow \text{CCH--TS7} \rightarrow \text{CCH--P1} \\[6pt]

\text{CCH--TS2} \rightarrow \text{CCH--M7}
\rightarrow \text{CCH--TS8} \rightarrow \text{CCH--P3}
\end{array}
\right.
\]

In addition, a deeply bound cyclic product well (CCH--P3) is identified,
which becomes accessible through the high-lying ring-closure transition
state CCH--TS8.
This feature highlights an \emph{additional bonding manifold} enabled by
C--C bond formation in the CCH + H$_2$CS system (including ring closure).
By contrast, the CN + H$_2$CS reaction can also explore a rich set of
approach geometries (including C-end vs.\ N-end attack), but does
not open an analogous C--C ring-closure channel.

\begin{table}[htbp]
\centering
\caption{
Relative energies (in kcal~mol$^{-1}$) of stationary points involved in the
CCH + H$_2$CS reaction.
Energies are given relative to the separated reactants CCH + H$_2$CS.
The first column corresponds to CCSD(T)-F12a+ZPE//DSD-PBEP86, and the second to
DSD-PBEP86+ZPE.
}
\label{tab:hcch_energetics}
\begin{tabular}{l c c}
\hline\hline
Species &
$\Delta E_{\mathrm{CCSD(T)\mbox{-}F12a+ZPE//DSD-PBEP86}}$ &
$\Delta E_{\mathrm{DSD-PBEP86+ZPE}}$ \\
\hline
CCH + H$_2$CS (reactants) & 0.00 & 0.00 \\
CCH--vdW                 & --   & -0.41 \\
CCH--TS0                 &--    & 0.12 \\
CCH--TS1                 & -49.6  & -51.27 \\
CCH--TS2                 & -42.5  & -33.15\\
CCH--TS3                 & -34.9  & -37.75 \\
CCH--TS4                 & -52.2  & -55.56\\
CCH--TS5                &  -79.6& -82.1\\
CCH--TS6                 & -36.9  & -40.48 \\
CCH--TS7                 & -32.9  & -32.73 \\
CCH--TS8                 & 16.6   & -2.97 \\
CCH--TS9                 & -39.5  & -41.01 \\
CCH--M1                  & -69.5  & -71.28 \\
CCH--M2                  & -80.7  & -86.05 \\
CCH--M3                  & -83.8  & -86.53\\
CCH--M4                  & -84.4  & -87.13 \\
CCH--M5                  & -90.3  & -94.15\\
CCH--M6                  & -71.7  & -71.31 \\
CCH--M7                  & -67.2  & -68.81 \\
CCH--P1 (HCSCCH + H)     & -36.3  & -40.72 \\
CCH--P2 (HCCH + HCS)     & -51.4  & -51.92 \\
CCH--P3 (cyclic adduct)  & -89.6  & -91.22 \\
\hline\hline
\end{tabular}
\end{table}

\subsection{Energetic Comparison}
The resulting DSD-PBEP86+ZPE and CCSD(T)-F12a+ZPE//DSD-PBEP86 relative energies are summarized in
Tables~\ref{tab:cn_energetics} and~\ref{tab:hcch_energetics}.
Overall, the standalone DSD-PBEP86+ZPE energetics are systematically more
stabilizing than the CCSD(T)-F12a+ZPE//DSD-PBEP86 composite values: most wells
and transition states are shifted downward by about 5--10~kcal~mol$^{-1}$
(median $\sim$5.4), while the long-range CN--vdW complex is essentially
unchanged. The largest deviations occur for the deeper addition well CN--M2
($\sim$10~kcal~mol$^{-1}$) and for a few bottlenecks (e.g., CN--TS7 and
CN--TS12), whereas the overall ordering of stationary points and product
exothermicities remains qualitatively consistent.

For CCH + H$_2$CS, DSD-PBEP86+ZPE is only slightly more stabilizing than the
CCSD(T)-F12a+ZPE//DSD-PBEP86 composite energetics, typically by
$\sim$1--4~kcal~mol$^{-1}$. The main discrepancies are confined to a few
bottlenecks: the emerged barrier CCH--TS8 shifts from +16.6 to
--2.97~kcal~mol$^{-1}$, and CCH--TS2 is $\sim$9~kcal~mol$^{-1}$ less
stabilized at the DSD-PBEP86 level. These differences mainly affect selected
barrier heights rather than the relative stability of the product channels.

\subsection{Entrance-Channel Capture and Kinetic Treatment}

\label{sec:capture_C6}
 
To parameterize the capture step and the entrance branching used in the
master-equation treatment, relaxed approach scans were computed along the
intermolecular separation coordinate \(R\). These scans show that the different
orientations of attack do not probe the same long-range interaction and
therefore cannot be represented by a single isotropic capture coefficient.
Accordingly, the entrance channels were described in MESS using effective,
orientation-dependent capture parameters.

Because the entrance scans involve open-shell fragments and weakly interacting
asymptotic states, special care was taken to avoid artifacts associated with
state switching or inconsistent reference states along the scan coordinate. The
scan geometries were first generated at the DSD-PBEP86/aug-cc-pVTZ level, and
the electronic-state character was monitored along \(R\) before the final
single-point refinements. The entrance potentials used for the capture fits
were then obtained from CCSD(T)-F12a single-point energies computed on these
scan geometries.

\begin{figure}[h!]
  \centering
  \includegraphics[width=0.72\linewidth]{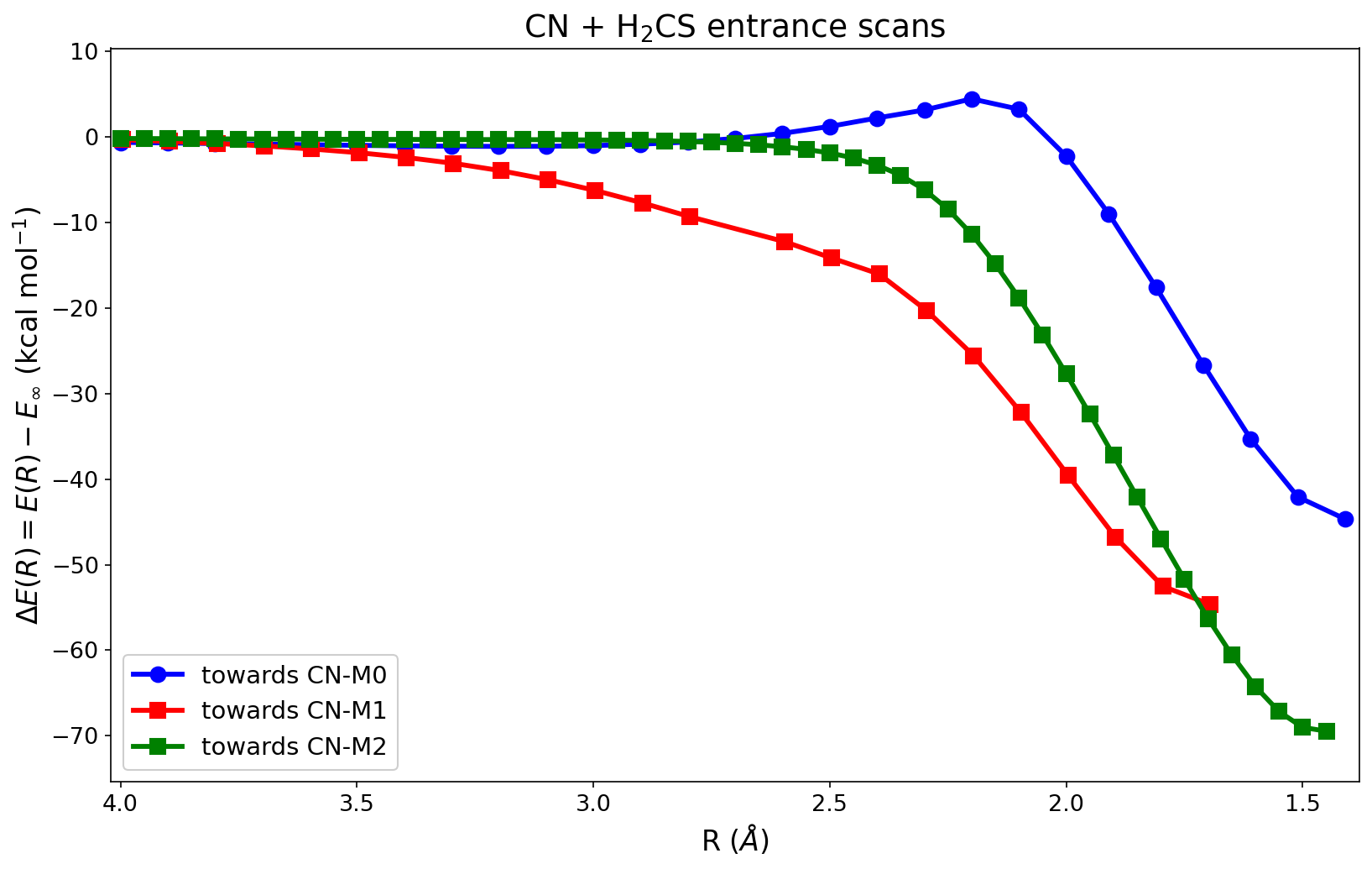}
  \caption{Short-range entrance-channel CCSD(T)-F12a potential scans for the CN + H$_2$CS system (R $\le$ 4~\AA), shown relative to the long-range asymptote, $\Delta E(R)=E(R)-E_\infty$, in kcal\,mol$^{-1}$. The three orientations leading toward the CN--vdW, CN--M1, and CN--M2 entrance
regions are displayed with distinct symbols/colors.}
  \label{fig:CN_H2CS_shortR_scans}
\end{figure}

\begin{figure}[h!]
  \centering
  \includegraphics[width=0.72\linewidth]{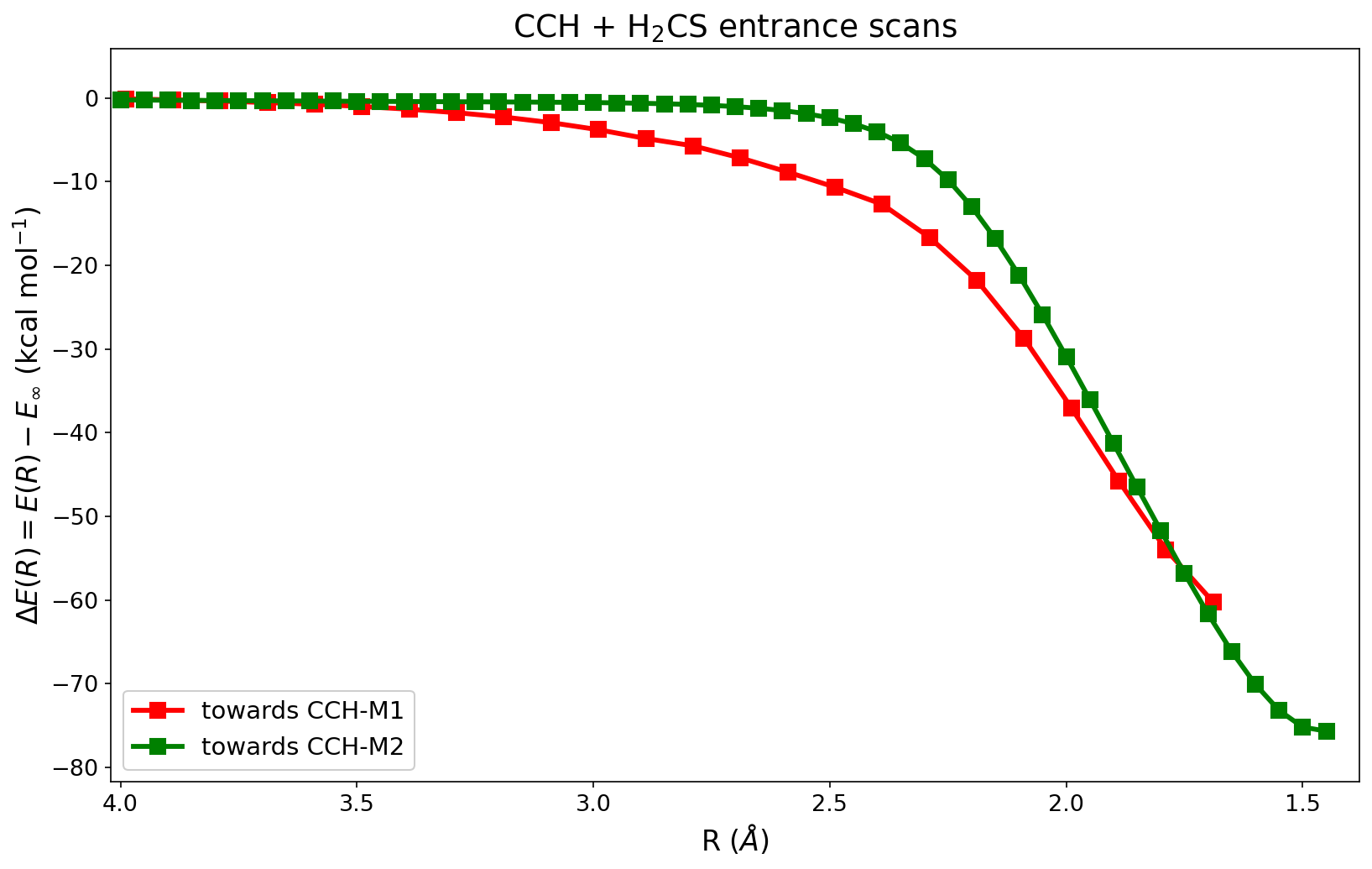}
  \caption{Short-range entrance-channel CCSD(T)-F12a potential scans for the CCH + H$_2$CS system (R $\le$ 4~\AA), shown relative to the long-range asymptote, $\Delta E(R)=E(R)-E_\infty$, in kcal\,mol$^{-1}$. The two orientations leading toward the CCH--M1 and CCH--M2 wells are displayed (red and green squares, respectively).}
  \label{fig:CCH_H2CS_shortR_scans}
\end{figure}

The short-range portions of the resulting CCSD(T)-F12a//DSD-PBEP86 entrance
potentials for CN + H$_2$CS and CCH + H$_2$CS are shown in Figs.~\ref{fig:CN_H2CS_shortR_scans} and~\ref{fig:CCH_H2CS_shortR_scans},
respectively, as \(\Delta E(R)=E(R)-E_\infty\) relative to the long-range
asymptote. Here, \(R\) denotes the scanned interfragment distance defining the
approach coordinate for each entrance channel. For the CN + H$_2$CS system,
\(R\) corresponds to the C--N distance for the CN--M0 approach, to the S--C
distance for the CN--M1 approach, and to the C--C distance for the CN--M2
approach.

In the following kinetic schemes, R1 denotes the separated bimolecular
reactants, namely R1 = CN + H$_2$CS for the CN system and
R1 = CCH + H$_2$CS for the CCH system.

For the CN + H\(_2\)CS system, three distinct entrance routes were retained.
Two of them correspond to barrierless oriented approaches toward the chemically
bound entrance adducts M1 and M2, while the third proceeds through a shallow
van der Waals entrance complex followed by the inner transition state TS0
toward M0:
\begin{equation}
\mathrm{R1}
\rightarrow
\begin{cases}
\mathrm{M1}, \\[2pt]
\mathrm{M2}, \\[2pt]
\mathrm{vdW}\rightarrow \mathrm{TS0}\rightarrow \mathrm{M0}.
\end{cases}
\end{equation}
The subsequent chemistry from M1, M2, and M0 was then propagated through the
connected multiwell network leading to the product channels.

For the CCH + H$_2$CS system, two oriented barrierless entrances were retained,
corresponding to the approaches toward M1 and M2,
\begin{equation}
\mathrm{R1}
\rightarrow
\begin{cases}
\mathrm{M1}, \\[2pt]
\mathrm{M2}.
\end{cases}
\end{equation}
The downstream network includes the product channels P1, P2, and P3. However,
the P3 channel remains kinetically negligible compared with P1 and P2 over the
computed temperature range.
 
For the capture treatment in MESS, the long-range part of each entrance
potential was described by an effective inverse-power dispersion form,
\begin{equation}
E(R)=E_\infty-\frac{C_6}{R^6},
\end{equation}
where \(R\) is expressed in bohr and \(C_6\) in atomic units
\((E_{\rm h}a_0^6)\). In order to obtain reliable asymptotic references, the
entrance scans were extended up to \(R=15~\text{\AA}\) at the CCSD(T)-F12a level
whenever possible. Each potential was shifted with respect to the largest-\(R\)
energy,
\begin{equation}
\Delta E(R)=E(R)-E_\infty,
\end{equation}
and the attractive tail was fitted to the \(-C_6/R^6\) form. Points affected by
short-range chemical interaction or by small numerical asymptotic noise were not
included in the final long-range fit.

The resulting \(C_6\) values should be regarded as effective,
orientation-dependent capture parameters rather than isotropic molecular
dispersion coefficients. They characterize the long-range attraction along the
specific approach coordinate leading to a given entrance channel. Therefore,
they are not expected to scale directly with the final well depths: a pathway
leading to a deep short-range adduct may still exhibit a weak long-range
attraction along the scanned coordinate.

For the CN + H\(_2\)CS system, the fitted effective coefficients were
\begin{align}
C_6(\mathrm{CN\!-\!vdW}) &= 560~E_{\rm h}a_0^6,\\
C_6(\mathrm{CN\!-\!M1})  &= 970~E_{\rm h}a_0^6,\\
C_6(\mathrm{CN\!-\!M2})  &= 60~E_{\rm h}a_0^6.
\end{align}
Here, CN--vdW denotes the shallow van der Waals entrance complex preceding
CN--TS0, while CN--M1 and CN--M2 correspond to the oriented barrierless
approaches toward the chemically bound entrance adducts. The ordering of the
\(C_6\) values reflects the anisotropy observed in the entrance scans: the
CN--M1 approach displays the strongest long-range attraction, CN--vdW represents
a moderate vdW capture pathway, whereas the CN--M2 scan remains nearly flat at
long range and becomes strongly stabilizing only at shorter intermolecular
separations.

For the CCH + H\(_2\)CS system, the corresponding effective capture parameters
were
\begin{align}
C_6(\mathrm{CCH\!-\!M1}) &= 340~E_{\rm h}a_0^6,\\
C_6(\mathrm{CCH\!-\!M2}) &= 160~E_{\rm h}a_0^6.
\end{align}
These channel-specific coefficients were used in the MESS input to describe the
orientation-dependent capture into the different entrance wells.
\subsection{Phenomenological Rate Coefficients and Product Branching}

At \(P=10^{-7}\)~atm, the phenomenological rate coefficients are governed
primarily by entrance capture followed by competition within the downstream
multiwell networks. Because the different approach geometries are associated
with distinct effective long-range \(C_6\) coefficients, the product-forming
rates were computed separately for each entrance description rather than by
forcing all incoming fluxes into a single global MESS graph.

For the CN + H$_2$CS system, separate calculations were performed for the
M1- and M2-initiated addition networks, as well as for the
vdW/TS0/M0-mediated pathway leading to P3. For the CCH + H$_2$CS system, the
two entrance calculations correspond to the M1- and M2-initiated networks. The
resulting rates should therefore be interpreted as entrance-specific
phenomenological contributions. For a given entrance channel \(i\), the
product-specific rate may be viewed schematically as
\begin{equation}
k_{i\rightarrow P_j}(T)=a_i(T)\,Y_{i\rightarrow P_j}(T),
\end{equation}
where \(a_i(T)\) is the entrance capture rate and
\(Y_{i\rightarrow P_j}(T)\) is the conditional yield toward product \(P_j\)
after propagation through the downstream network.

For CN + H$_2$CS, the M1-initiated rates are larger than the corresponding
M2-initiated rates, consistently with the stronger effective long-range capture
coefficient for the M1 approach,
\(C_6(\mathrm{CN\!-\!M1})=970~E_{\rm h}a_0^6\), compared with
\(C_6(\mathrm{CN\!-\!M2})=60~E_{\rm h}a_0^6\). This difference mainly reflects
anisotropic entrance capture rather than a downstream kinetic bottleneck.

\begin{figure}[h!]
  \centering
  \includegraphics[width=0.9\linewidth]{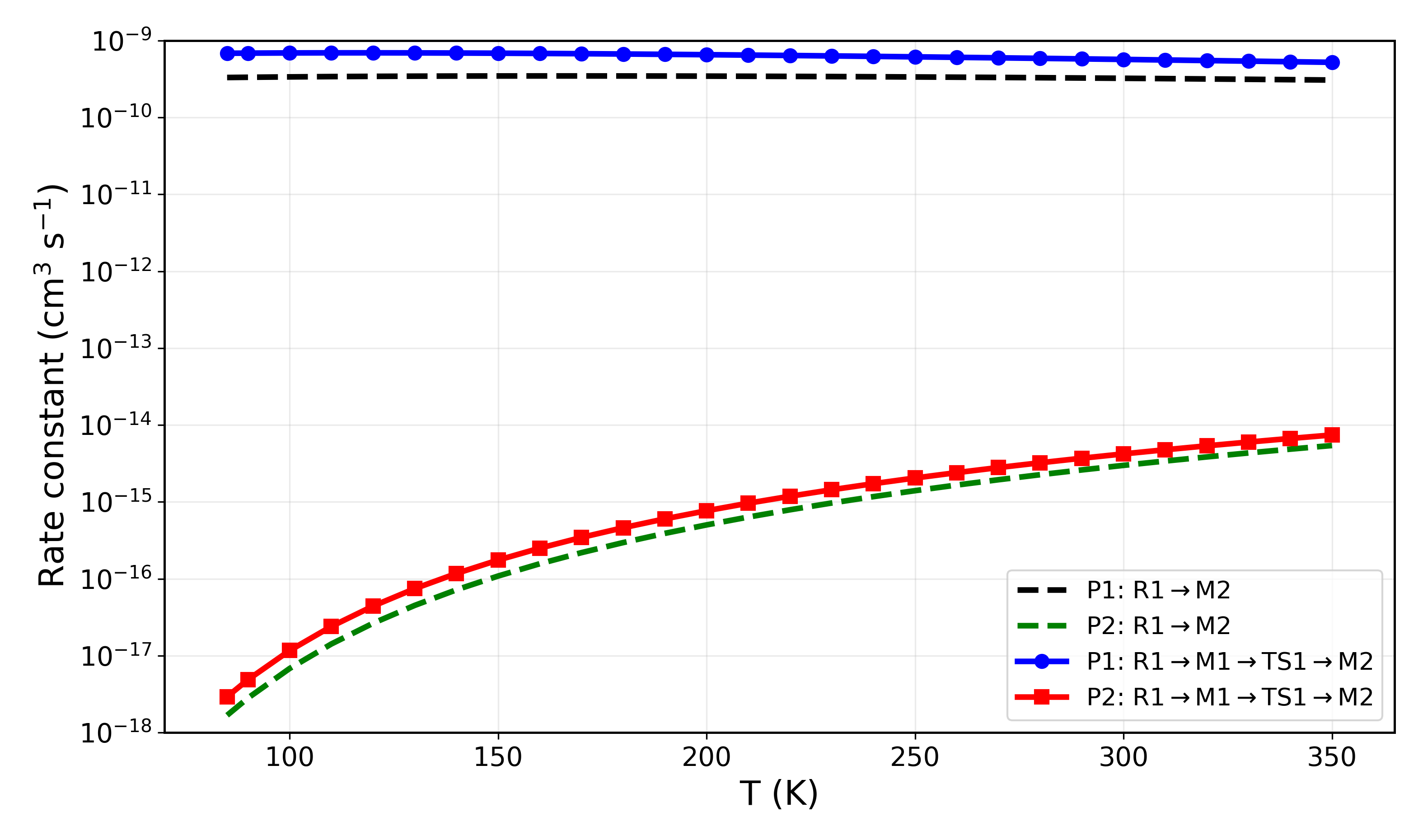}
  \caption{
  Temperature-dependent phenomenological rate coefficients for the
  CN + H$_2$CS reaction at a pressure of $P = 10^{-7}$~atm, comparing
  two entrance-network treatments. Solid lines correspond to the extended
  network initiated through R1$\rightarrow$M1$\rightarrow$TS1$\rightarrow$M2,
  whereas dashed lines correspond to the simplified network starting
  directly from R1$\rightarrow$M2. Blue/black curves denote the
  R1$\rightarrow$P1 (HCSCN + H) channel, whereas red/green curves denote the
  R1$\rightarrow$P2 (HCN + HCS) channel.
  }
  \label{fig:rates_CN_H2CS}
\end{figure}

The CN + H$_2$CS rate coefficients displayed in
Fig.~\ref{fig:rates_CN_H2CS} show a strong product selectivity toward
HCSCN + H (CN--P1). The HCN + HCS (CN--P2) channel remains several orders of
magnitude smaller, while the vdW/TS0/M0-mediated HNC + HCS (CN--P3) route makes
only a minor contribution over the explored temperature range.

\begin{figure}[h!]
  \centering
  \includegraphics[width=0.9\linewidth]{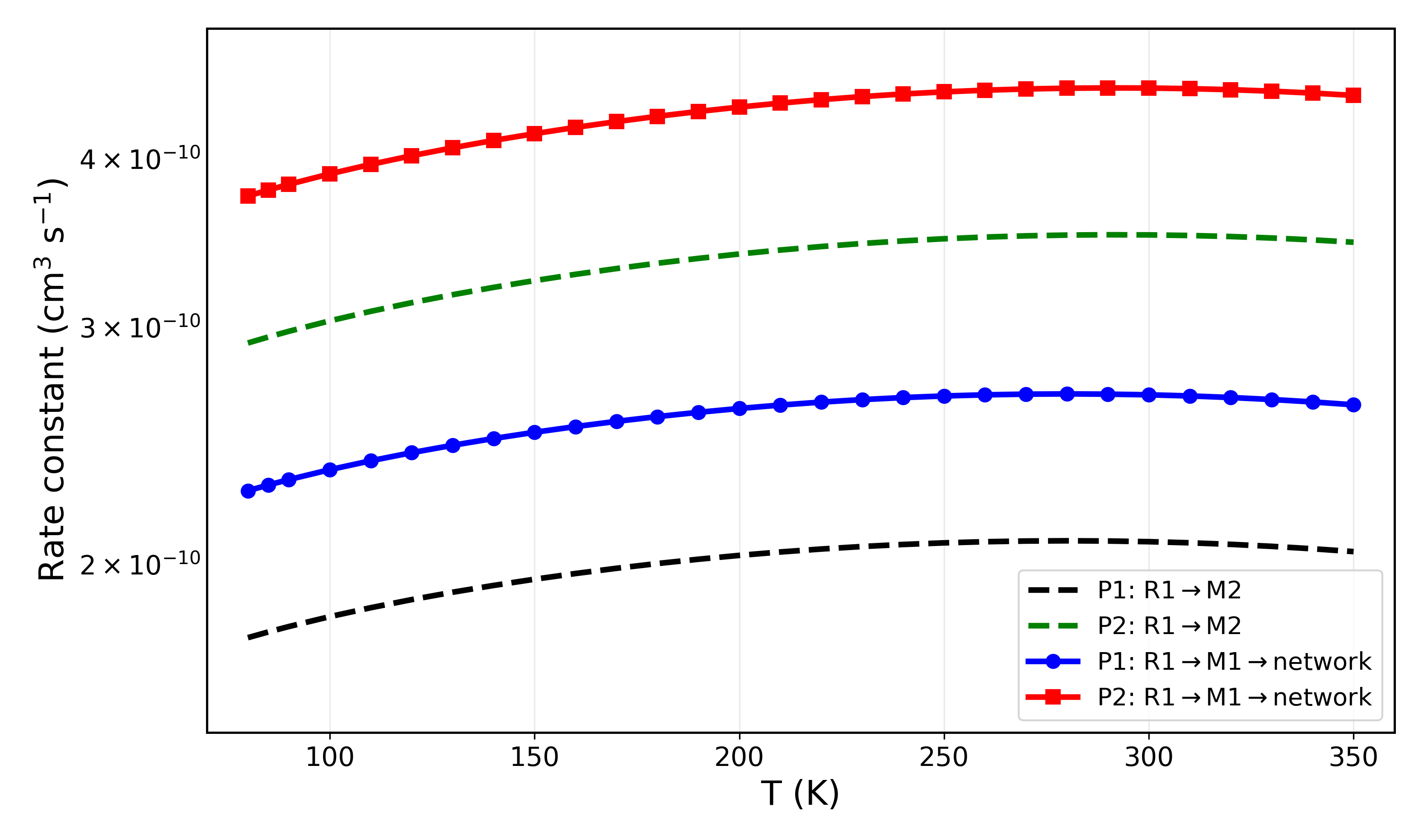}
  \caption{
  Temperature-dependent phenomenological rate coefficients for the
  CCH + H$_2$CS reaction at a pressure of $P = 10^{-7}$~atm, comparing
  two entrance-network treatments. Solid lines correspond to the extended
  network initiated through R1$\rightarrow$M1$\rightarrow$TS1$\rightarrow$M2,
  whereas dashed lines correspond to the simplified network starting
  directly from R1$\rightarrow$M2. Blue/black curves denote the
  R1$\rightarrow$P1 channel (HCSCCH + H), and red/green curves denote the
  R1$\rightarrow$P2 channel (HCCH + HCS).
  }
  \label{fig:rates_CCH_H2CS}
\end{figure}

For CCH + H$_2$CS, the two entrance capture coefficients are closer,
\(C_6(\mathrm{CCH\!-\!M1})=340~E_{\rm h}a_0^6\) and
\(C_6(\mathrm{CCH\!-\!M2})=160~E_{\rm h}a_0^6\), and the relative P1/P2
branching is similar for the two entrance descriptions
(Fig.~\ref{fig:rates_CCH_H2CS}). In contrast to the CN system, the CCH reaction
does not channel the flux almost exclusively into the detected product
HCSCCH + H (CCH--P1). Instead, HCCH + HCS (CCH--P2) carries the larger
fraction of the reactive flux, while HCSCCH + H remains a substantial secondary
channel.

Within the numerically converged temperature ranges considered here, the
capture-controlled product-forming rates display only a weak temperature
dependence, consistent with barrierless long-range association. For a pure
\(-C_6/R^6\) interaction, an approximate \(k_{\rm cap}\propto T^{1/6}\)
dependence provides a useful qualitative reference.

The CN + H$_2$CS results can be compared with the recent kinetic study of
Alessandrini et al. \citep{Alessandrini2025}. Both studies agree that
HCSCN + H is the dominant CN-driven product channel, but the absolute rates
differ substantially. Alessandrini et al. reported Kooij rate coefficients for
the HC(S)CN + H channel of the order of
$2.4\times10^{-11}$~cm$^3$~s$^{-1}$ at 100 K and
$6.5\times10^{-12}$~cm$^3$~s$^{-1}$ at 200 K, whereas the present
entrance-resolved treatment gives total HCSCN-forming rates of the order of
\(8\times10^{-10}\) to \(1.0\times10^{-9}\) cm$^3$~s$^{-1}$ over the explored
temperature range. This difference most likely reflects the treatment of the
entrance region: Alessandrini et al. used an isotropic \(C_6/R^6\) capture
description fitted along a single interfragment coordinate, whereas the present
work separates several orientation-dependent entrance channels.

The numerical values of the product-forming rate coefficients for the
CN + H$_2$CS and CCH + H$_2$CS reactions are reported in
Appendix~\ref{app:rate_tables}, Tables~\ref{tab:cn_h2cs_rates}
and~\ref{tab:cch_h2cs_rates}, respectively.

For use in astrochemical modeling, the entrance-specific product-forming rate
coefficients were also fitted to the modified Arrhenius, or Kooij, expression
\begin{equation}
k(T)=A\left(\frac{T}{300}\right)^{\beta}
\exp\!\left(-\frac{\gamma}{T}\right),
\label{eq:kooij}
\end{equation}
where \(T\) is in K and \(k(T)\) is in cm$^3$~s$^{-1}$. Fits were performed in
log-space over the numerically converged temperature range available for each
entrance-specific dataset. The resulting parameters are reported in
Table~\ref{tab:kooij_fits}. These analytical expressions are intended as compact
representations of the direct MESS rate coefficients, whereas the mechanistic
and astrochemical discussion below is based on the computed master-equation
rates.

\begin{table}[H]
\centering
\caption{Kooij fit parameters for the entrance-specific product-forming
phenomenological rate coefficients obtained from Eq.~\ref{eq:kooij}. The rate
coefficients are in cm$^3$~s$^{-1}$ and the temperature is in K.}
\label{tab:kooij_fits}
\begin{tabular}{c l c c c c}
\hline\hline
System & Channel & \(A\) & \(\beta\) & \(\gamma\) (K) & RMS$_{\log_{10}}$ \\
\hline
CN  & \(\mathrm{R1}\rightarrow\mathrm{M1}\rightarrow\mathrm{P1}\) 
    & \(7.413\times10^{-10}\) & \(-0.6696\) & 80.17 & 0.0042 \\

CN  & \(\mathrm{R1}\rightarrow\mathrm{M1}\rightarrow\mathrm{P2}\) 
    & \(3.220\times10^{-14}\) & 1.6667 & 610.06 & 0.0052 \\

CN  & \(\mathrm{R1}\rightarrow\mathrm{M2}\rightarrow\mathrm{P1}\) 
    & \(3.890\times10^{-10}\) & \(-0.3853\) & 56.28 & 0.0036 \\

CN  & \(\mathrm{R1}\rightarrow\mathrm{M2}\rightarrow\mathrm{P2}\) 
    & \(2.196\times10^{-14}\) & 1.9132 & 598.48 & 0.0047 \\

CN  & \(\mathrm{R1}\rightarrow\mathrm{vdW}\rightarrow\mathrm{P3}\) 
    & \(5.684\times10^{-14}\) & 1.2871 & 3850.10 & 0.0061 \\

CCH & \(\mathrm{R1}\rightarrow\mathrm{M1}\rightarrow\mathrm{P1}\) 
    & \(2.895\times10^{-10}\) & \(-0.0415\) & 24.85 & 0.0036 \\

CCH & \(\mathrm{R1}\rightarrow\mathrm{M1}\rightarrow\mathrm{P2}\) 
    & \(4.851\times10^{-10}\) & \(-0.0092\) & 22.38 & 0.0034 \\

CCH & \(\mathrm{R1}\rightarrow\mathrm{M2}\rightarrow\mathrm{P1}\) 
    & \(2.252\times10^{-10}\) & \(-0.0411\) & 24.81 & 0.0036 \\

CCH & \(\mathrm{R1}\rightarrow\mathrm{M2}\rightarrow\mathrm{P2}\) 
    & \(3.773\times10^{-10}\) & \(-0.0088\) & 22.34 & 0.0034 \\
\hline\hline
\end{tabular}
\end{table}

\section{Astrochemical Implications}
\subsection{Kinetic Diagnostics for HCSCN and HCSCCH Formation}

To compare the two detected sulfur-bearing species on a common kinetic basis,
we define the relative kinetic indicator
\begin{equation}
\label{eq:Q_ratio}
Q(T) \equiv 
\frac{k_{\mathrm{CN+H_2CS}\rightarrow \mathrm{HCSCN+H}}(T)}
     {k_{\mathrm{CCH+H_2CS}\rightarrow \mathrm{HCSCCH+H}}(T)} .
\end{equation}
The numerator is the total HCSCN-forming rate coefficient obtained by summing
the P1 contributions from the M1- and M2-initiated CN entrance calculations,
whereas the denominator is the corresponding total HCSCCH-forming P1 rate
coefficient obtained from the M1- and M2-initiated CCH calculations.

\begin{figure}[h!]
  \centering
  \includegraphics[width=0.90\linewidth]{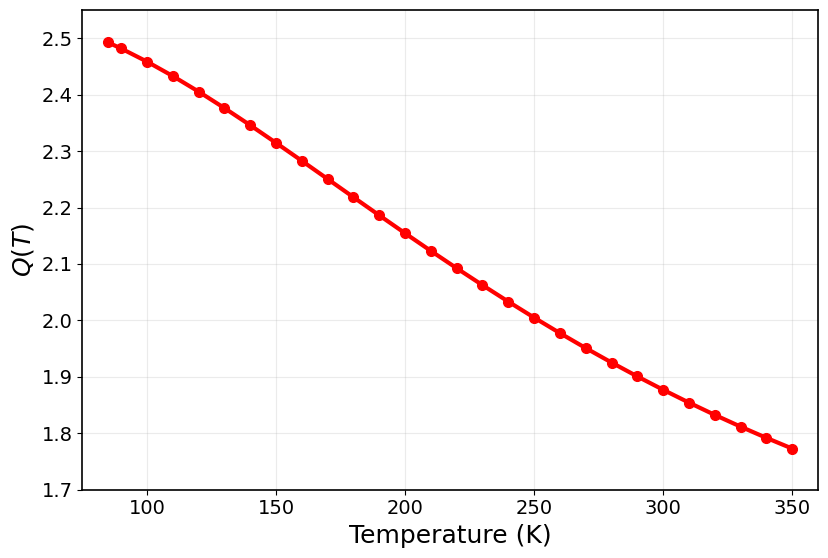}
  \caption{Temperature dependence of the relative kinetic indicator $Q(T)$,
  defined in Eq.~\ref{eq:Q_ratio}, comparing the HCSCN-forming rate from
  CN + H$_2$CS with the HCSCCH-forming rate from CCH + H$_2$CS, as derived from
  master-equation calculations at $P=10^{-7}$~atm.}
  \label{fig:Q_ratio}
\end{figure}

Over the common 85--350~K temperature range, \(Q(T)\) remains larger than
unity, decreasing from about 2.49 at 85~K to about 1.77 at 350~K
(Fig.~\ref{fig:Q_ratio}). Thus, within the restricted gas-phase reaction
subset studied here, the HCSCN-forming rate is larger than the
HCSCCH-forming rate. This ratio should not be interpreted as an abundance
ratio, but rather as a kinetic indicator for the two product-forming routes.

\begin{figure}[h!]
  \centering
  \includegraphics[width=0.90\linewidth]{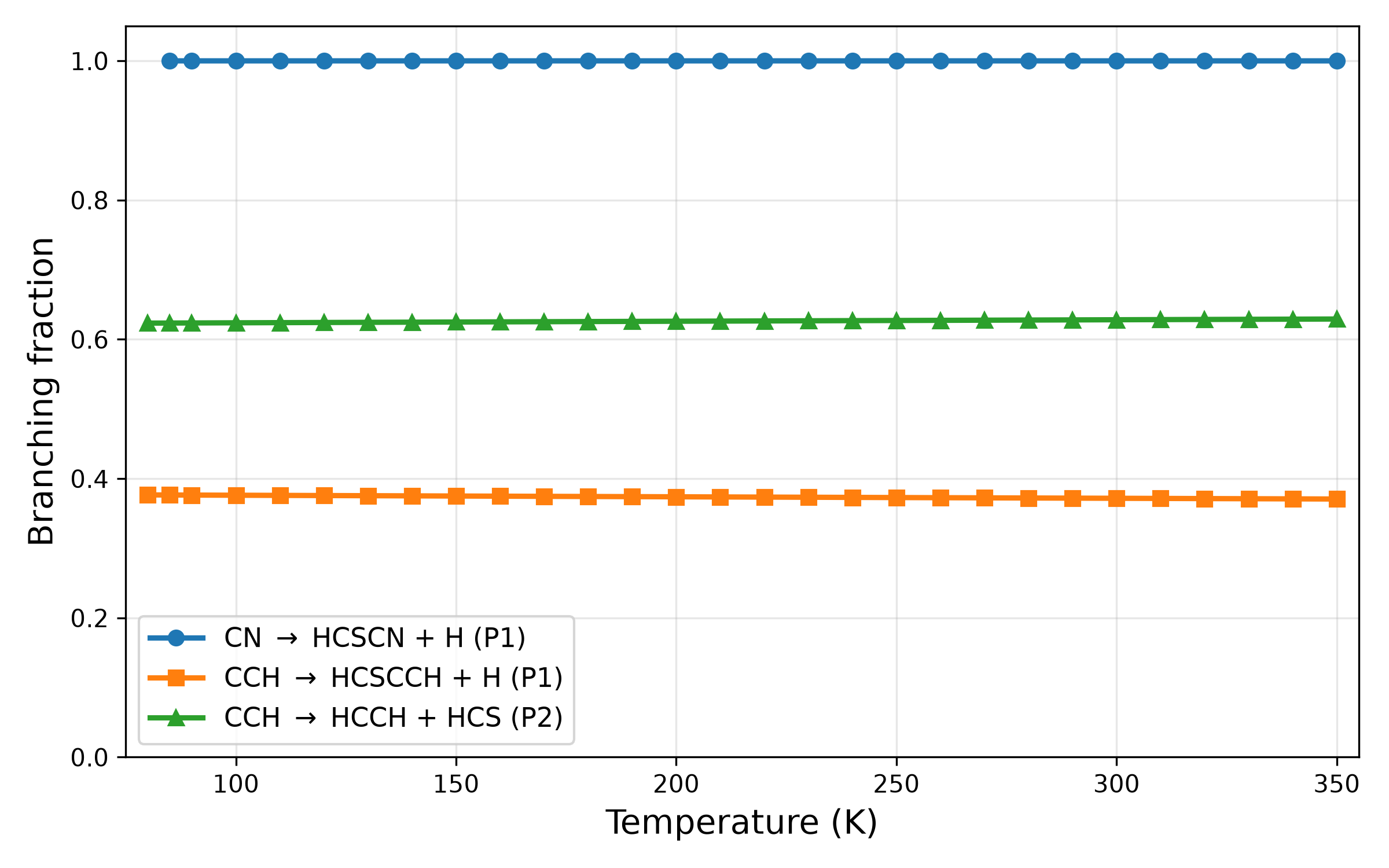}
  \caption{
  Temperature dependence of the summed product branching fractions for the
  dominant product channels of the CN + H$_2$CS and CCH + H$_2$CS reactions at
  \(P=10^{-7}\)~atm. The CN reaction is almost exclusively selective toward
  HCSCN + H (P1), whereas the CCH reaction distributes the reactive flux mainly
  between HCSCCH + H (P1) and HCCH + HCS (P2). Minor channels with negligible
  branching fractions are not shown.
  }
  \label{fig:branching_CN_CCH}
\end{figure}

The summed product branching fractions shown in
Fig.~\ref{fig:branching_CN_CCH} provide a complementary view. For
CN + H$_2$CS, the branching is overwhelmingly dominated by HCSCN + H, while
the minor HCN + HCS and HNC + HCS channels are negligible on the scale of the
dominant product. For CCH + H$_2$CS, the reactive flux is shared mainly between
HCSCCH + H and HCCH + HCS, with the latter channel carrying the larger
fraction of the flux. The CCH branching fraction toward HCCH + HCS increases
slightly from about 0.623 at 85~K to about 0.629 at 350~K, whereas the
HCSCCH + H fraction decreases from about 0.377 to about 0.371. The explicit
definitions used for the summed rates and branching fractions are given in
Appendix~\ref{app:branching_definitions}.

These diagnostics suggest that CN + H$_2$CS can efficiently feed HCSCN under
gas-phase conditions, whereas CCH + H$_2$CS does not map uniquely onto HCSCCH
formation because a larger fraction of the reactive flux is diverted toward
HCCH + HCS. All else being equal, this provides a mechanistic contribution to
the different observed behavior of HCSCN and HCSCCH in sources such as TMC-1.
A quantitative comparison with abundances, however, requires a full
astrochemical network including the CN/CCH ratio, the availability of H$_2$CS,
competing formation and destruction pathways, updated gas-phase networks, and
grain-surface or ice-mantle chemistry
\citep{Millar1997,Boogert2015,Boogert2022,NatComm2022_Sulfur,ACSESC2024,NatComm2025_Sulfur}.

A useful comparison can be made with the recently investigated
C$_2$H + CH$_2$O reaction. Douglas et al. \citep{Douglas2024_C2H_CH2O}
combined low-temperature experiments with MESMER calculations and found that
H-abstraction to form C$_2$H$_2$ + CHO dominates, whereas C- and O-addition
channels contribute less than 0.3\% even at 600 K. This contrasts with the
present CCH + H$_2$CS system, where sulfur substitution leads to barrierless
entrance-channel addition and deep intermediate wells, making HCSCCH + H a
chemically viable, although non-dominant, product channel.

Although TMC-1 is a \(\sim 10\) K source, the present MESS calculations are
reported only over the numerically converged temperature range: 80--350 K for
CCH + H$_2$CS and 85--350 K for CN + H$_2$CS.

\begin{figure}[h!]
  \centering
  \includegraphics[width=0.90\linewidth]{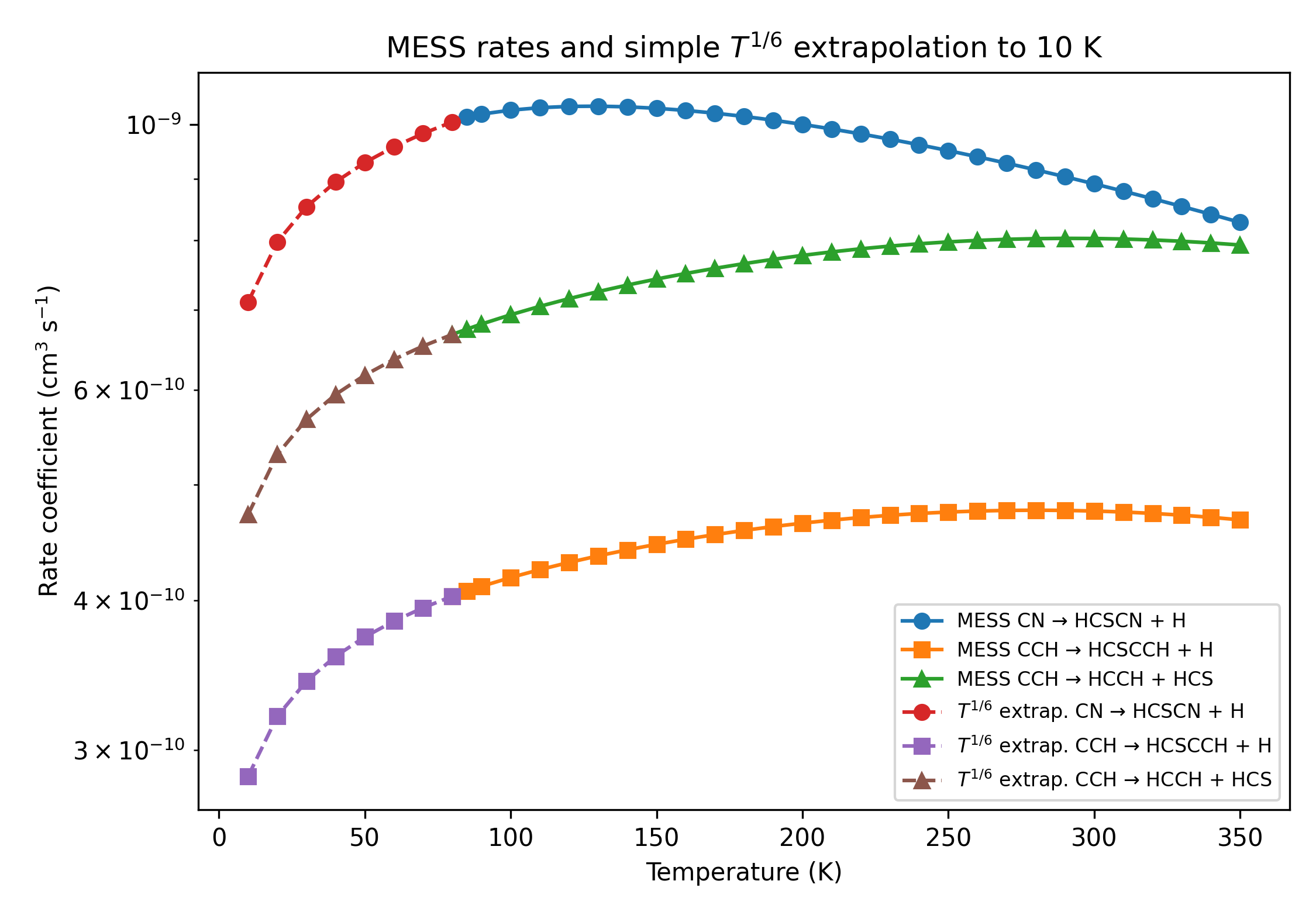}
  \caption{
  Comparison between the MESS product-forming rate coefficients and a simple
  low-temperature capture-law extrapolation, \(k(T)\propto T^{1/6}\), down to
  10 K. Solid curves correspond to the summed MESS rates over the numerically
  converged temperature range. Dashed curves show the extrapolation from the
  lowest available MESS temperature: 85 K for CN + H$_2$CS and 80 K for
  CCH + H$_2$CS.  
  }
  \label{fig:T16_extrapolation}
\end{figure}

A simple capture-law extrapolation, \(k(T)\propto T^{1/6}\), was nevertheless
performed from the lowest numerically converged temperatures to assess whether
the qualitative ordering of the dominant channels would be preserved down to
10 K. As shown in Fig.~\ref{fig:T16_extrapolation}, the extrapolated rates
remain of the same order of magnitude, and the CN-driven HCSCN-forming channel
remains more efficient than the CCH-driven HCSCCH-forming channel. This extrapolation was not used in the Kooij fitting or in the quantitative
branching analysis.

\FloatBarrier
\section{Conclusion}

We have presented an \textit{ab initio} and master-equation kinetic study of
the CN + H$_2$CS and CCH + H$_2$CS reactions, motivated by the detection of
HCSCN and HCSCCH in cold interstellar environments. Both systems proceed
through barrierless entrance capture into deep chemically bound wells, but they
display markedly different product selectivities.

For CN + H$_2$CS, the reaction is strongly selective toward HCSCN + H. The
dominant flux proceeds through the M1/M2 addition network, whereas the competing
HCN + HCS and HNC + HCS channels remain negligible over the computed temperature
range. The additional vdW/TS0/M0-mediated route to HNC + HCS contributes only
weakly under the low-pressure conditions considered here.

For CCH + H$_2$CS, the product distribution is less selective. HCSCCH + H is a
chemically viable product channel, but HCCH + HCS carries the larger fraction
of the reactive flux, accounting for approximately 62--63\% of the
product-forming flux, while HCSCCH + H contributes approximately 37--38\% over
the 80--350 K range. The cyclic P3 channel is energetically stable but
kinetically negligible because of the high-lying ring-closure transition state.

The entrance-channel analysis shows that the absolute rate coefficients are
sensitive to the orientation-dependent long-range capture treatment. The M1-
and M2-initiated calculations therefore reflect distinct entrance geometries
and effective \(C_6\) coefficients rather than a simple downstream kinetic
bottleneck. More generally, the present results show that product
exothermicity alone is insufficient to predict molecular formation
efficiencies: entrance topology, capture anisotropy, and multiwell kinetic
competition control the final branching ratios.

These results identify CN + H$_2$CS as a selective gas-phase route to HCSCN,
whereas CCH + H$_2$CS provides a viable but non-dominant route to HCSCCH.
Future work will extend this gas-phase picture to ice-coated interstellar
grains, where adsorption, diffusion, and reactive processing may further modify
the effective formation rates and branching ratios of sulfur-bearing carbon
chains \citep{Molpeceres2026_CN_IcyGrains}.

\appendix
\section{Product-Forming Rate Coefficients}
\label{app:rate_tables}

\begin{table}[H]
\centering
\caption{Product-forming rate coefficients (cm$^3$~s$^{-1}$) at $P=10^{-7}$~atm for CN + H$_2$CS.}
\label{tab:cn_h2cs_rates}
\begin{tabular}{c c c c}
\hline\hline
$T$ (K) & $k_{\mathrm{P1}}$ & $k_{\mathrm{P2}}$ & $k_{\mathrm{P3}}$ \\
\hline
85 & 1.01501e-09 & 4.60649e-18 & 2.31732e-34 \\
90 & 1.02005e-09 & 7.69154e-18 & 3.14607e-33 \\
100 & 1.02776e-09 & 1.84987e-17 & 2.65608e-31 \\
110 & 1.03270e-09 & 3.82375e-17 & 1.00409e-29 \\
120 & 1.03524e-09 & 7.06050e-17 & 2.08009e-28 \\
130 & 1.03570e-09 & 1.19612e-16 & 2.71633e-27 \\
140 & 1.03437e-09 & 1.89365e-16 & 2.46780e-26 \\
150 & 1.03149e-09 & 2.83966e-16 & 1.67792e-25 \\
160 & 1.02726e-09 & 4.07646e-16 & 9.02285e-25 \\
170 & 1.02186e-09 & 5.64352e-16 & 3.99657e-24 \\
180 & 1.01544e-09 & 7.57785e-16 & 1.50667e-23 \\
190 & 1.00813e-09 & 9.92325e-16 & 4.96037e-23 \\
200 & 1.00004e-09 & 1.27096e-15 & 1.45510e-22 \\
210 & 9.91269e-10 & 1.59748e-15 & 3.86646e-22 \\
220 & 9.81887e-10 & 1.97535e-15 & 9.43262e-22 \\
230 & 9.71972e-10 & 2.40780e-15 & 2.13654e-21 \\
240 & 9.61580e-10 & 2.89775e-15 & 4.53432e-21 \\
250 & 9.50763e-10 & 3.44776e-15 & 9.08566e-21 \\
260 & 9.39574e-10 & 4.06169e-15 & 1.73082e-20 \\
270 & 9.28055e-10 & 4.74139e-15 & 3.15070e-20 \\
280 & 9.16242e-10 & 5.48995e-15 & 5.50928e-20 \\
290 & 9.04176e-10 & 6.30949e-15 & 9.28984e-20 \\
300 & 8.91888e-10 & 7.20243e-15 & 1.51624e-19 \\
310 & 8.79411e-10 & 8.17069e-15 & 2.40209e-19 \\
320 & 8.66773e-10 & 9.21669e-15 & 3.70613e-19 \\
330 & 8.54005e-10 & 1.03421e-14 & 5.57836e-19 \\
340 & 8.41130e-10 & 1.15486e-14 & 8.21044e-19 \\
350 & 8.28171e-10 & 1.28373e-14 & 1.18416e-18 \\
\hline\hline
\end{tabular}
\end{table}

\begin{table}[H]
\centering
\caption{Product-forming rate coefficients (cm$^3$~s$^{-1}$) at $P=10^{-7}$~atm for CCH + H$_2$CS.}
\label{tab:cch_h2cs_rates}
\begin{tabular}{c c c c}
\hline\hline
$T$ (K) & $k_{\mathrm{P1}}$ & $k_{\mathrm{P2}}$ & $k_{\mathrm{P3}}$ \\
\hline
80  & 4.03270e-10 & 6.67437e-10 & 3.63440e-61 \\
85  & 4.07239e-10 & 6.74335e-10 & 1.55980e-58 \\
90  & 4.11010e-10 & 6.80913e-10 & 3.40031e-56 \\
100 & 4.18031e-10 & 6.93220e-10 & 3.19563e-52 \\
110 & 4.24455e-10 & 7.04556e-10 & 5.65699e-49 \\
120 & 4.30371e-10 & 7.15068e-10 & 2.86762e-46 \\
130 & 4.35846e-10 & 7.24863e-10 & 5.56527e-44 \\
140 & 4.40924e-10 & 7.34012e-10 & 5.07915e-42 \\
150 & 4.45638e-10 & 7.42567e-10 & 2.53720e-40 \\
160 & 4.50006e-10 & 7.50562e-10 & 7.76964e-39 \\
170 & 4.54040e-10 & 7.58009e-10 & 1.59190e-37 \\
180 & 4.57747e-10 & 7.64920e-10 & 2.33058e-36 \\
190 & 4.61124e-10 & 7.71290e-10 & 2.57532e-35 \\
200 & 4.64169e-10 & 7.77112e-10 & 2.23681e-34 \\
210 & 4.66877e-10 & 7.82377e-10 & 1.58251e-33 \\
220 & 4.69241e-10 & 7.87071e-10 & 9.36868e-33 \\
230 & 4.71253e-10 & 7.91180e-10 & 4.75113e-32 \\
240 & 4.72909e-10 & 7.94694e-10 & 2.10384e-31 \\
250 & 4.74202e-10 & 7.97602e-10 & 8.26906e-31 \\
260 & 4.75128e-10 & 7.99894e-10 & 2.92386e-30 \\
270 & 4.75686e-10 & 8.01565e-10 & 9.40620e-30 \\
280 & 4.75874e-10 & 8.02613e-10 & 2.78138e-29 \\
290 & 4.75695e-10 & 8.03038e-10 & 7.62366e-29 \\
300 & 4.75148e-10 & 8.02839e-10 & 1.95082e-28 \\
310 & 4.74240e-10 & 8.02029e-10 & 4.69274e-28 \\
320 & 4.72977e-10 & 8.00610e-10 & 1.06701e-27 \\
330 & 4.71365e-10 & 7.98595e-10 & 2.30567e-27 \\
340 & 4.69414e-10 & 7.95999e-10 & 4.75223e-27 \\
350 & 4.67133e-10 & 7.92835e-10 & 9.38066e-27 \\
\hline\hline
\end{tabular}
\end{table}
 
\section{Explicit Definitions of Product Branching Fractions}
\label{app:branching_definitions}

For the branching fractions shown in Fig.~\ref{fig:branching_CN_CCH}, the
entrance-specific product-forming rates were first summed explicitly for each
product channel.

For the CN + H$_2$CS reaction, the total product-forming rates were defined as
\begin{align}
K_{\mathrm{CN,P1}}(T) &=
k_{\mathrm{CN,M1}\rightarrow \mathrm{P1}}(T)
+
k_{\mathrm{CN,M2}\rightarrow \mathrm{P1}}(T),\\
K_{\mathrm{CN,P2}}(T) &=
k_{\mathrm{CN,M1}\rightarrow \mathrm{P2}}(T)
+
k_{\mathrm{CN,M2}\rightarrow \mathrm{P2}}(T),\\
K_{\mathrm{CN,P3}}(T) &=
k_{\mathrm{CN,vdW/M0}\rightarrow \mathrm{P3}}(T).
\end{align}
Here, P1, P2, and P3 correspond to HCSCN + H, HCN + HCS, and HNC + HCS,
respectively. The CN product branching fractions were then obtained as
\begin{equation}
\mathrm{BF}_{\mathrm{CN,Pn}}(T)=
\frac{K_{\mathrm{CN,Pn}}(T)}
{K_{\mathrm{CN,P1}}(T)+K_{\mathrm{CN,P2}}(T)+K_{\mathrm{CN,P3}}(T)}
\qquad (n=1,2,3).
\end{equation}

For the CCH + H$_2$CS reaction, the total product-forming rates were defined as
\begin{align}
K_{\mathrm{CCH,P1}}(T) &=
k_{\mathrm{CCH,M1}\rightarrow \mathrm{P1}}(T)
+
k_{\mathrm{CCH,M2}\rightarrow \mathrm{P1}}(T),\\
K_{\mathrm{CCH,P2}}(T) &=
k_{\mathrm{CCH,M1}\rightarrow \mathrm{P2}}(T)
+
k_{\mathrm{CCH,M2}\rightarrow \mathrm{P2}}(T).
\end{align}
Here, P1 and P2 correspond to HCSCCH + H and HCCH + HCS, respectively. The
cyclic P3 channel was included in the kinetic network but remains negligible
over the computed temperature range. The CCH product branching fractions shown
in Fig.~\ref{fig:branching_CN_CCH} were therefore obtained as
\begin{equation}
\mathrm{BF}_{\mathrm{CCH,Pn}}(T)=
\frac{K_{\mathrm{CCH,Pn}}(T)}
{K_{\mathrm{CCH,P1}}(T)+K_{\mathrm{CCH,P2}}(T)}
\qquad (n=1,2).
\end{equation}
 
\section*{Acknowledgements}
This publication is based upon work from COST Action CA21126, Carbon Molecular Nanostructures in Space (NanoSpace), supported by COST (European Cooperation in Science and Technology). The authors would like to acknowledge the Deanship of Graduate Studies and Scientific Research, Ta\"if University, Saudi Arabia, for funding this work. This work was granted access to the HPC resources of MesoPSL financed by the Région Île-de-France and the project Equip@Meso (reference ANR-10-EQPX-29-01) of the programme Investissements d’Avenir supervised by the Agence Nationale de la Recherche.
\subsection*{Supporting Information}
The data that support the findings of this study are available on request from the corresponding author.

 \subsection*{Notes}
The authors declare no competing financial interest.

\bibliography{references}
 \end{document}